# Seamlessly merging radar ranging/imaging, wireless communications, and spectrum sensing, for 6G empowered by microwave photonics

Taixia Shi[a,†], Yang Chen[a,†,*], and Jianping Yao[b,*]

[a] Shanghai Key Laboratory of Multidimensional Information Processing, School of Communication and Electronic Engineering, East China Normal University, Shanghai 200241, China
[b] Microwave Photonic Research Laboratory, School of Electrical Engineering and Computer Science, University of Ottawa, Ottawa, ON K1N 6N5, Canada
[†]These authors contributed equally to this work.
*Correspondence to: Y. Chen, ychen@ce.ecnu.edu.cn; J. Yao, jpyao@uottawa.ca.

**ABSTRACT**
Integration of radar, wireless communications, and spectrum sensing is being investigated for 6G with an increased spectral efficiency. Microwave photonics (MWP), a technique that combines microwave engineering and photonic technology to take advantage of the wide bandwidth offered by photonics for microwave signal generation and processing is considered an effective solution for the implementation of the integration. In this paper, an MWP-assisted joint radar, wireless communications, and spectrum sensing (JRCSS) system that enables precise perception of the surrounding physical and electromagnetic environments while maintaining high-speed data communication is proposed and demonstrated. Communication signals and frequency-sweep signals are merged in the optical domain to achieve high-speed radar ranging and imaging, high-data-rate wireless communications, and wideband spectrum sensing. In an experimental demonstration, a JRCSS system supporting radar ranging with a measurement error within ±4 cm, two-dimensional imaging with a resolution of 25×24.7 mm, wireless communications with a data rate of 2 Gbaud, and spectrum sensing with a frequency measurement error within ±10 MHz in a 6-GHz bandwidth, is demonstrated.

## 1. Introduction
In addition to high-speed wireless communications, to realize intelligent interconnection between people, machines, and things, various sensing functions and artificial intelligence will also be incorporated into a 6G network to provide digital infrastructure for new types of applications [1], [2]. Figure 1 shows a scenario in which multi-functions including target sensing, spectrum sensing, and wireless communications are incorporated in a 6G network. Instead of using independent systems for multiple functions, one highly recommended and cost-effective solution is to integrate multidimensional functions into a single system, to unify microwave ranging and imaging, wireless communications, and spectrum sensing functions. This integrated system is known as a joint radar, wireless communications, and spectrum sensing (JRCSS) system, which has already been investigated in the electrical domain [3], [4]. However, conventional electronic solutions have encountered challenges in addressing the bottleneck of broadband high-frequency complex waveform generation and have struggled to meet the high data rate and high-resolution application requirements due to the limited tunability and reconfigurability of electronic systems. Microwave photonics [5], [6], a technique that combines microwave engineering and photonic technology, has received significant attention due to its numerous advantages, including large bandwidth, high frequency, low loss, wide tunability, and immunity to electromagnetic interference. Over the past few decades, microwave photonics has been extensively explored for the development of advanced radar [7], wireless communications [8], spectrum sensing [9], and multifunctional integrated systems [10], [11].

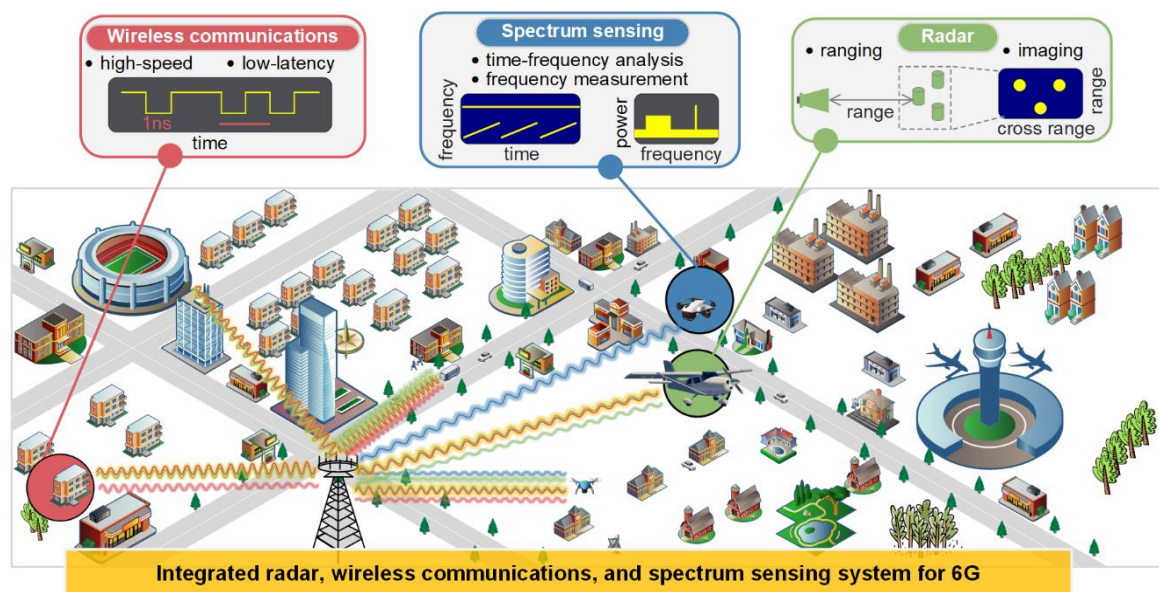

Fig. 1 Application scenario of the integrated radar, wireless communications, and spectrum sensing system.

Thanks to the benefits offered by microwave photonics, this technology is widely regarded as a promising solution for 6G systems, and the main research hotspots are the generation of millimeter-wave or terahertz wireless communication signals with high data rates using photonics-assisted signal generation methods [12], [13], and the realization of low-loss, long-distance wireless signal distribution using wireless-over-fiber techniques [14], [15]. As the advancements in microwave photonics continue to drive innovation, there is a growing recognition of its potential impact on various sectors, fostering developments in radar systems and spectrum sensing techniques. This has the potential to revolutionize industries and pave the way for seamlessly merging radar ranging/imaging, spectrum sensing, and communications functions. The first all-optical implementation of microwave photonic radar was reported in ref. [16], and the report of this work has driven the rapid development of research on microwave photonic radar systems. In the follow-up work [17]–[19], people try to integrate the generation of broadband linearly frequency-modulated (LFM) waveforms with the echo de-chirping process in the optical domain to achieve a high range and imaging resolution. Spectrum sensing powered by microwave photonics can achieve frequency measurement [20]–[24] and time-frequency analysis [25]–[30] by mapping the frequency information of the signal under test (SUT) to power [20], [21], space [21], or time [22]–[30]. The spectrum sensing via frequency-to-power mapping [20] is commonly not suitable for multi-frequency measurement without being combined with other methods, while the spectrum sensing via frequency-to-space mapping is commonly achieved by channelization and always needs to be combined with other methods [21]. Spectrum sensing that relies on frequency-to-time mapping (FTTM) has the ability of multi-frequency measurement and can be effectively implemented by utilizing fiber dispersion [24]–[28] or by sweeping optical signals in combination with optical filtering [22], [23], [29], [30].

Building upon the aforementioned investigation of single-function microwave photonic systems, researchers have subsequently extended their studies to explore multi-function systems. These include joint communications and radar (JCR) systems, as well as joint radar and spectrum sensing systems, both benefiting from advanced microwave photonic technology. The radar signal and communication signal can be frequency-division [31], [32] and time-division [33] multiplexed in the JCR system. To share the same waveform in the JCR system, quadrature phase-shift keying (QPSK) signal [34] and spectrum-spreading phase-coding signal [35] were used. Nevertheless, a high sampling rate is required in the radar receivers in refs [34], [35]. Amplitude-shift keying LFM (ASK-LFM) signal [36] and QPSK-LFM signal [37], [38] were used in the photonics-assisted JCR system in conjunction with the de-chirping operation, successfully reducing the complexity of the radar receiver. In the photonics-assisted joint radar and spectrum sensing system, an LFM signal is commonly used for radar detection and spectrum sensing, where the de-chirping operation is used for further radar signal processing and the original LFM signal is used for spectrum sensing via FTTM. The broadband LFM signal can be obtained by photonic frequency doubling [39], photonic frequency quadrupling [40], or optically injection [41].

However, microwave photonic systems that can simultaneously realize the aforementioned three functions, which are highly desired in 6G, have not been studied. To bridge this gap, for the first time, we introduce a photonics-enabled approach that seamlessly integrates radar ranging/imaging, wireless communications, and spectrum sensing within a unified framework. An ASK communication signal and an LFM signal are converted to the optical domain, in which the LFM part of the modulated optical signal is used for spectrum sensing and the LFM part and ASK part of the modulated optical signal are fused after photodetection for radar and communication functions. An experiment is performed to verify the concept. A JRCSS system supporting radar ranging with a measurement error within ±4 cm, two-dimensional imaging with a resolution of 25×24.7 mm, wireless communications with a data rate of 2 Gbaud, and spectrum sensing with a frequency measurement error within ±10 MHz in a 6-GHz bandwidth, is demonstrated.

## 2. Principle

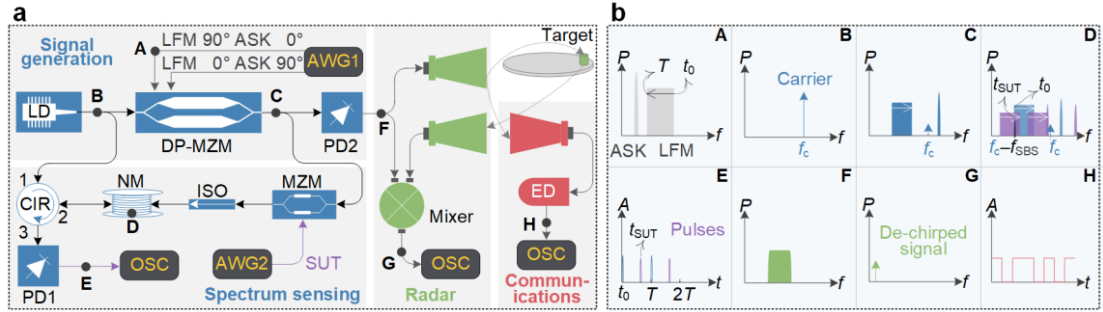

Fig. 2 Schematic of the proposed JRCSS system. **a** System structure diagram. **b** Spectra and waveforms at different locations (A–H) in **a**. LD, laser diode; DP-MZM, dual-parallel Mach–Zehnder modulator; LFM, linearly frequency-modulated signal; ASK, amplitude-shift keying signal; AWG, arbitrary waveform generator; PD, photodetector; MZM, Mach–Zehnder modulator; ISO, isolator; NM, nonlinear medium; CIR, circulator; ED, envelope detector; OSC, oscilloscope.

The schematic of the proposed JRCSS system is shown in Fig. 2a, and Fig. 2b shows the schematic diagrams of the signals at locations A–H in the system diagram in Fig. 2a. A continuous-wave optical wave generated by a laser diode (LD) is split into two parts. One part from the LD is injected into a dual-parallel Mach–Zehnder modulator (DP-MZM) to which an intermediate frequency (IF) ASK signal and a negative chirped LFM signal are applied. Note that the ASK signal and the negative chirped LFM signal have two different phase relationships with one being 90º and 0º and the other being 0º and 90º, as shown in Fig. 2a. Through carrier-suppressed tandem single-sideband (CS-TSSB) modulation at the DP-MZM, a double sideband signal with the IF ASK signal and the LFM signal located at the opposite sides of the optical carrier is generated, as shown in Fig. 2b C.

For spectrum sensing, one part of the optical signal from the DP-MZM is sent to a Mach–Zehnder modulator (MZM), to which an SUT is applied. By controlling the bias voltage, carrier-suppressed double-sideband (CS-DSB) modulation with a carrier suppression ratio to make the carrier at a power level close to its sideband is generated. The not fully suppressed optical carrier is used to provide a reference for frequency identification to simplify the operation. At the MZM, the LFM sideband of the CS-TSSB-modulated signal from the DP-MZM is employed as a frequency-sweep carrier. Therefore, after CS-DSB modulation, a series of frequency-sweep optical sidebands determined by the SUT frequency components will appear together with the partially suppressed optical carrier. The optical signal at the output of the MZM serves as a probe wave and is sent into a nonlinear medium (NM) for triggering the stimulated Brillouin scattering (SBS) effect. The other part of the optical signal from the LD functions as a pump wave and is sent into the NM from the other side through an optical circulator (CIR). The pump wave then interacts with the counter-propagating frequency-sweeping probe wave and generates an SBS gain, which serves as a narrowband optical bandpass filter. If the SBS gain spectrum is within the frequency-sweep ranges of the frequency-sweep optical carrier and SUT sidebands, the frequency-sweep optical carrier and SUT sidebands will be filtered by the SBS gain spectrum at a specific time in a single sweep period, as shown in Fig. 2b D. The specific time associated with the SUT sidebands is determined by the SUT frequency, while that linked to the frequency-sweep optical carrier not fully suppressed is a constant and used as a reference. After SBS interaction, optical pulses are generated at the specific time, and FTTM is implemented. Subsequently, the optical signal from the optical CIR is sent to a photodetector (PD1), where the optical pulses are converted into electrical pulses, as shown in Fig. 2b E. The electrical pulses are monitored by an oscilloscope (OSC), enabling the determination of the SUT frequency based on the time of pulse appearance. It is assumed that the carrier frequency of the LD is $f_c$, the center frequency of the SBS gain is $f_c - f_{SBS}$, the frequency-sweep range and period of the frequency-sweep optical signal are $f_c - f_{SBS} \sim f_c - f_{SBS} + f_B$ and $T$. Here, $f_B$ is the frequency-sweep bandwidth of the frequency-sweep optical signal, and $f_{SBS}$ is the Brillouin frequency shift (BFS). Under these circumstances, the spectrum sensing range is $0 \sim f_B$. After FTTM, if the reference pulse appears at $t_0$ and the signal pulse appears at $t_{SUT}$, the SUT frequency can be obtained by $f_{SUT} = f_B \times (t_{SUT} - t_0)/T$. Besides the SUT frequency, the time-frequency information of the SUT can also be obtained under high-speed frequency sweeping conditions by combining the pulses obtained from different sweep periods [29]. Therefore, spectrum sensing is realized in the system using the optical LFM sideband generated from the DP-MZM.

For radar ranging and imaging, the other part of the optical signal from the DP-MZM is injected into a second PD (PD2) to generate an electrical ASK-LFM signal by beating the LFM sideband and the ASK sideband. One part of the electrical signal from PD2 is sent to a transmitting antenna, which radiates the ASK-LFM signal into the free space. The radar-receiving antenna captures the echo signal reflected by the target, which is then sent to an RF mixer via the RF port. The other part of the electrical signal from PD2 is sent to the mixer via the local oscillator (LO) port. After mixing at the mixer, a de-chirped signal is generated from the IF port of the mixer, which is captured by an OSC and processed to enable target ranging and inverse synthetic aperture radar (ISAR) imaging.

For communications, the communication-receiving antenna receives the ASK-LFM signal radiated from the transmitting antenna. The signal envelope of the ASK signal is then obtained through an envelope detector (ED), and the original communication data can be obtained using a decision circuit. To do so, the received signal from the communication-receiving antenna is split into two by an electrical coupler (EC), and the two output signals from the EC are fed into a mixer via the RF port and LO port through two RF cables with an identical length. After self-mixing and low-pass filtering, the signal envelope carrying the communication data can also be obtained.

## 3. Experiment and results

3.1 Experimental setup

An experiment is carried out based on the setup shown in Fig. 2. A light wave, characterized by a wavelength of 1553.096 nm and a power of 16 dBm, is generated by an LD (ID Photonics CoBriteDX1-1-C-H01-FA) and subsequently divided into two equal parts by an optical coupler (OC1). One output of OC1 is sent to a DP-MZM (Fujitsu FTM7961EX). Two channels of AWG1 (Keysight M8195A, 64 GSa/s) are used to generate the signals applied to the DP-MZM for CS-TSSB modulation. To implement the CS-TSSB modulation, the negative chirped LFM signal and the IF ASK signal need to be combined with two different phase relationships, which can commonly be realized by using two 90° hybrid couplers and two 3-dB couplers to couple the negative chirped LFM signal and the IF ASK signal like the TSSB modulation [42]. To simplify the experiment, the two signals with different phase relationships after 90° hybrid couplers and two 3-dB couplers are directly generated by AWG1, as shown in Fig. 2. The two signals from AWG1 have a peak-to-peak amplitude of 1 V. The LFM signal has a center frequency of 7.8 GHz, a bandwidth of 6 GHz, and a period of 4 μs. The center frequency and baud rate of the ASK signal are set to 3 GHz and 0.5 or 2 Gbaud. The two signals from AWG1 are applied to the two RF ports of the DP-MZM, respectively, and the bias condition of the DP-MZM for the CS-TSSB modulation is the same as that for the carrier-suppressed single-sideband (CS-SSB) modulation [43]. Subsequently, the optical signal from the DP-MZM is amplified by an erbium-doped fiber amplifier (EDFA, Amonics AEDFA-PA-35-B-FA) and then equally split into two parts using OC2.

In one output of OC2 for spectrum sensing, the optical signal is CS-DSB modulated at an MZM (Fujitsu, FTM7938EZ) by the SUT generated from AWG2 (Keysight M8190A, 10 GSa/s) with a peak-to-peak amplitude of 1 V. The optical signal from the MZM is utilized as the probe wave and sent to an NM (a section of 25.2-km single-mode fiber), where it interacts with the counter-propagating pump wave from OC1. After SBS interaction, the SUT frequency is mapped to the time of occurrence of the optical pulse. Then, the optical pulses from the optical CIR are sent to PD1 (Nortel Networks PP-10G) to convert them to electrical pulses. The electrical pulses with SUT frequency information from PD1 are captured by the OSC (Rohde & Schwarz RTO2032) with a sampling rate is 100 MSa/s.

Another part of the optical signal from OC2 is injected into PD2 (u2t MPRV1331A) to generate an electrical ASK-LFM signal for radar and communication functions. The ASK-LFM signal from PD2 is amplified by an electrical amplifier (EA1, (ALM 145-5023-293 5.85 to 14.5 GHz, 23 dB) and then equally split by EC1 (Narda 4456-2, 2 to 18 GHz, −3 dB). One output of EC1 is injected into the LO port of Mixer1 (M/A-COM M14A). Another output of EC1 is emitted by a transmitting antenna (GHA080180-SMF-14, 8 to 18 GHz). The radar-receiving antenna (GHA080180-SMF-14, 8 to 18

GHz) captures the reflected signal from the target, which is then amplified by EA2 (CLM 145-7039-293B, 5.85 to 14.50 GHz, 39 dB) and sent to the RF port of Mixer1. The de-chirped signal from the IF port of Mixer1 is captured by the OSC with a sampling rate of 4 MSa/s and processed to enable target ranging and ISAR imaging.

The communication-receiving antenna (GHA080180-SMF-14, 8 to 18 GHz) receives the ASK-LFM signal from the transmitting antenna. The received ASK-LFM signal is amplified by EA3 (CLM 145-7039-293B, 5.85 to 14.50 GHz, 39 dB). The signal envelope is then obtained by self-mixing via EC2 (Narda 4456-2, 2 to 18 GHz, −3 dB) and Mixer2 (Miteq m30).

3.2 Signal generation

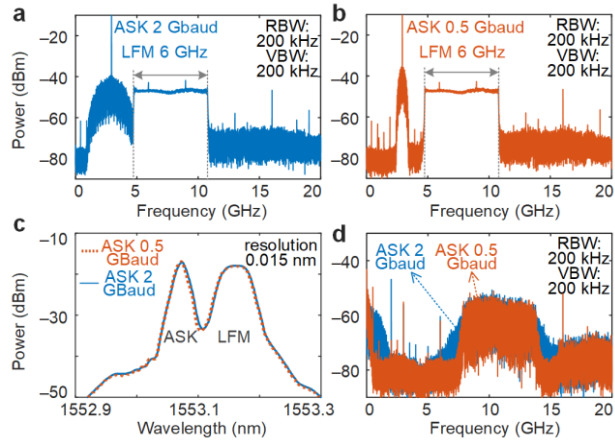

Fig. 3 Signal generation results. Electrical spectra of the signal from the AWG when the baud rate of the ASK signal is **a** 2 Gbaud and **b** 0.5 Gbaud. **c** Optical spectra of the optical signal from the DP-MZM. **d** Electrical spectra of the electrical signal from PD2.

An electrical signal with a peak-to-peak amplitude of 1 V is generated by an arbitrary waveform generator (AWG1). The electrical signal consists of an ASK signal and an LFM signal. Figure 3a shows the spectrum of the electrical signal with the ASK signal having a baud rate of 2 Gbaud and Fig. 3b shows the spectrum of the electrical signal with the ASK signal having a baud rate of 0.5 Gbaud. In the two cases, the LFM signal has a negative chirp and a frequency-sweep range of 4.8~10.8 GHz. The electrical signal is applied to the DP-MZM. Another electrical signal, also consisting of the same ASK signal and LFM signal but with a different phase relationship, is also generated by AWG1 and applied to the DP-MZM but from a different output channel.

The optical spectra of the LFM optical sideband and the ASK optical sideband at the output of the DP-MZM are shown in Fig. 3c. After CS-TSSB modulation, the ASK signal and the LFM signal are loaded to the opposite sides of the optical carrier due to the specially designed phase relationships. The spectra in the two cases are almost identical due to the limited resolution of 0.015 nm of the optical spectrum analyzer (OSA, ANDO AQ6317B) used to measure the optical spectrum. The LFM optical sideband is used for spectrum sensing while the beating product of the ASK optical sideband and LFM optical sideband is used for radar ranging, imaging, and communication functions.

Figure 3d shows the electrical spectra of the generated ASK-LFM signal from PD2 with a center frequency of 10.8 GHz. The power of this signal is measured to be −18.5 dBm. The spectrum of the ASK-LFM signal is broadened to a certain extent on the basis of the spectrum of the original LFM signal, and the degree of broadening is dependent on the baud rate of the ASK signal. In addition, baseband ASK signals, as well as some single-tone interferences, are also observed. The single-tone interferences are mainly from AWG and harmonics of the ASK signal carrier.

3.3 High-speed communications

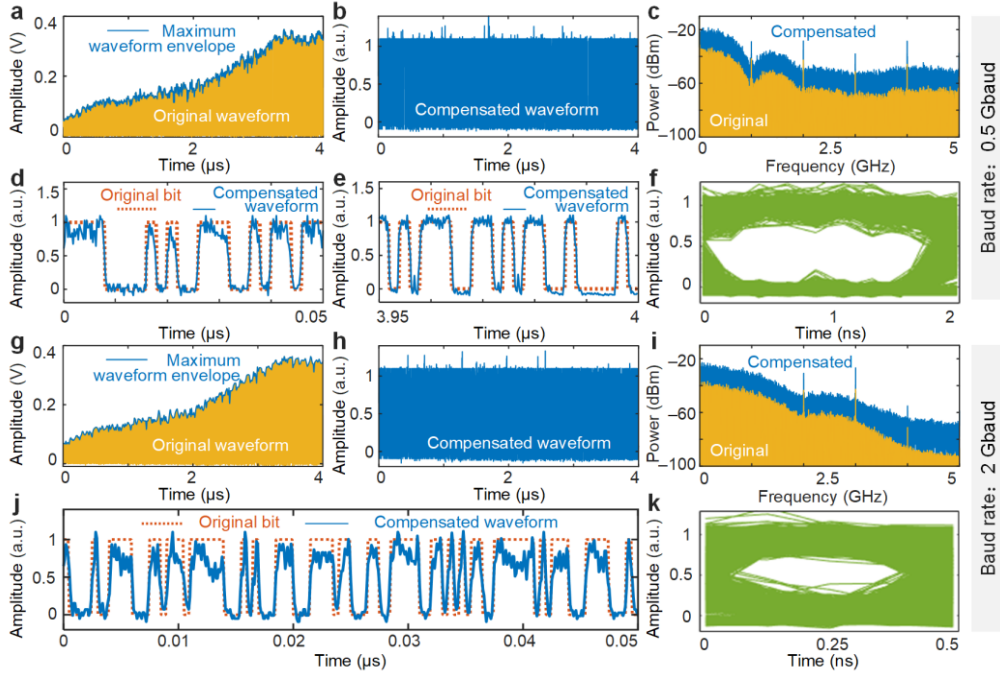

Fig. 4 High-speed communication results. **a-f** The baud rate is 0.5 Gbaud. **a** The original waveform captured by the OSC and **b** the compensated waveform. **c** Electrical spectra corresponding to **a** and **b**. A section of the compensated waveform from **d** 0 to 0.05 μs, and **e** 3.95 to 4 μs. **f** Eye diagrams of **b**. **g-k** The baud rate is 2 Gbaud. **g** The original waveform captured by the OSC and **h** the compensated waveform. **i** Electrical spectra corresponding to **g** and **h**. **j** A section of the compensated waveform from 0 to 0.05 μs. **k** Eye diagrams of **h**.

The two ASK-LFM signals generated above are first used for high-speed communications. Figure 4a shows the 4-μs temporal waveform of the self-mixing signal captured by the OSC with a sampling rate of 10 GSa/s when the baud rate of the ASK-LFM signal is 0.5 Gbaud. It should be noted that due to the limited bandwidth of OSC and its intrinsic low-pass filtering properties, the signal envelope can be obtained without additional low-pass filtering in the analog or digital domain. It can be seen, the amplitude of the waveform is noticeably uneven and gradually increases with time. Due to the uneven response of the mixer over a wide bandwidth, it exhibits a higher response at lower frequencies and a lower response at higher frequencies. Because a negative chirped LFM signal is used, the high-frequency signal appears first and is followed by the low-frequency signal within one period of the LFM signal, resulting in a gradual increase in amplitude of the signal after self-mixing.

To mitigate the impact of the uneven mixer frequency response, the temporal waveform envelope is extracted to compensate for the original waveform. After compensation, the waveform is flattened, as illustrated in Fig. 4b. Figure 4c displays the electrical spectra of the original waveform and the compensated waveform by fast Fourier transform (FFT). Figure 4d and e exhibit a segment of the compensated temporal waveform from 0 to 0.05 µs and from 3.95 to 4 µs, respectively. As can be seen, the original bit information is well represented by the compensated waveform after self-mixing, which indicates that the communication function of the system is well implemented. Figure 4f shows the eye diagrams of the compensated waveform, the eyes are widely open, confirming good receiving and detection of the communication signal.

Then, the baud rate is increased to 2 Gbaud. The frequency-sweep range of the LFM signal is the same as the case when the baud rate is 0.5 Gbaud. Therefore, the amplitude unevenness of the waveform in Fig. 4g is very similar to that in Fig. 4a. The compensated waveform is shown in Fig. 4h, with a section from 0 to 0.05 µs shown in Fig. 4j. Again, the original bit information is well recovered. The spectra of the waveform before and after compensation and the eye diagram are shown in Fig. 4i and k. The eye-opening in Fig. 4k is worse than that in Fig. 4f, mainly because of the increased data rate, which is four times that in Fig. 4f. For practical applications, direct envelope detection using an envelope detector or self-mixing using a mixer with a flatter response over the wide bandwidth can mitigate the uneven amplitude of the waveform and better results are expected.

3.4 Radar ranging and imaging.

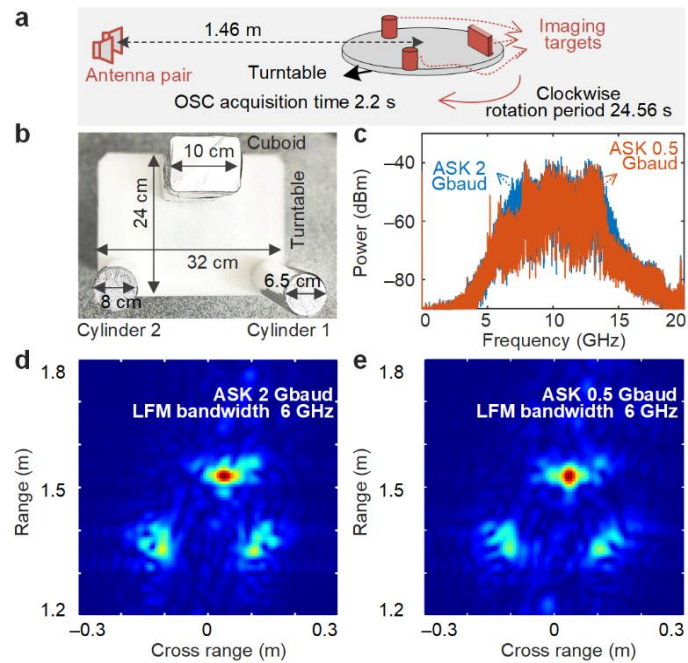

Fig. 5 Target imaging results. **a** Schematic for target imaging. **b** A cuboid and two cylinders used for target sensing. **c** Electrical spectra of the echo signal. Imaging result when the ASK baud rate is **d** 2 Gbaud and **e** 0.5 Gbaud.

The same ASK-LFM signal is then used for target radar ranging and imaging. Figure 5a shows the schematic for moving target imaging. To implement ISAR imaging, a cuboid and two cylinders covered with silver paper are placed on a turntable and employed as imaging targets, as shown in Fig. 5b. The dimensions of the cuboid are 10 cm (length)×8 cm (width)×18 cm (height). The diameter and height of cylinder 1 are 6.5 and 12 cm, whereas that of cylinder 2 are 8 and 10 cm. The range from the center of the turntable to the antenna pair is approximately 1.47 m. In the ISAR imaging experiment, the turntable is rotated clockwise with a period of 24.56 s. A de-chirped signal is captured by the OSC with an accumulation time of 2.2 seconds when the cuboid is farthest from the antenna, so the theoretical cross-range resolution is 24.7 mm [19], [44]. Since the LFM bandwidth is 6 GHz, the theoretical range resolution is 25 mm [19], [44]. When the turntable is stationary, the electrical spectra of the echo signal after amplification are shown in Fig. 5c, with the corresponding signal power of approximately −8.7 dBm. In Fig. 5c, ripples are observed in the electrical spectra of the echo signal, which is the result of overlapping multiple echo signals from the three targets. Figure 5d and e show the imaging results of the targets when the baud rates of the ASK-LFM signal are 2 Gbaud and 0.5 Gbaud, respectively. The three targets can be well distinguished. Because the function of target sensing is mainly determined by the LFM part of the ASK-LFM signal, there is no obvious difference in image quality at the two ASK baud rates.

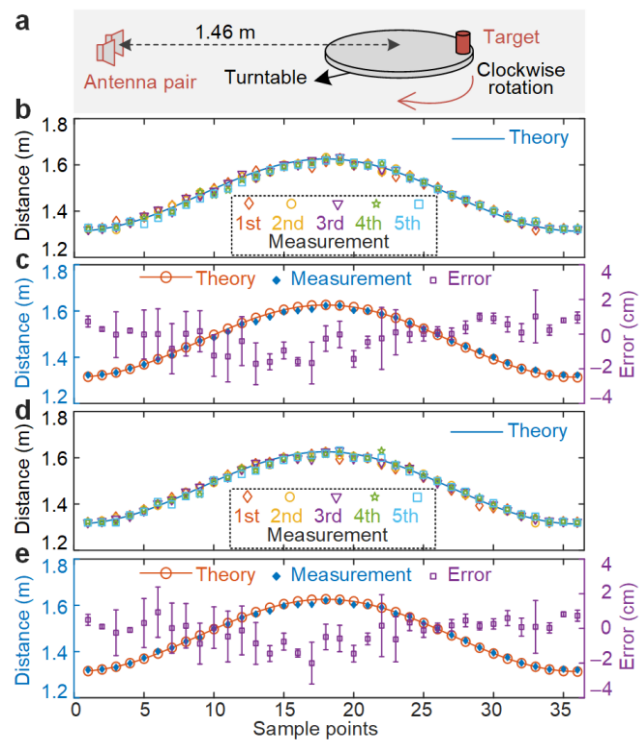

Fig. 6 Moving target ranging results. **a** Schematic for moving target ranging. **b** Five sets of ranging results and **c** average ranging results and the corresponding errors when the baud rate of the ASK-LFM signal is 0.5 Gbaud. **d** Five sets of ranging results and **e** average ranging results and the corresponding errors when the baud rate of the ASK-LFM signal is 2 Gbaud.

Figure 6a shows the schematic for target ranging when only a cylinder is placed on and rotated with the turntable. The range between the turntable and antenna pair is not changed. The rotation period of the turntable is also 24.56 s, during which the range between the cylinder and the antenna is continuously monitored. When the baud rate of the ASK-LFM signal is 0.5 Gbaud, the results at 35 equally spaced sampling points with a 4-μs sampling time in a rotation period are measured 5 times and given in Fig. 6b along with the theoretical curve. The measured target ranges are consistent with the theory. The red circles, blue diamonds, and purple error bars in Fig. 6c show the theoretical target range, the average range results of the five measurements in Fig. 6b, and the standard deviations of the range errors corresponding to Fig. 6b. The measurement error is no more than 4 cm. Figure 6d and e show the measurement results when the baud rate of the ASK-LFM signal is increased to 2 Gbaud, and similar performance can be obtained, which means the communication data rate in the ASK-LFM signal does not have a significant impact on target ranging.

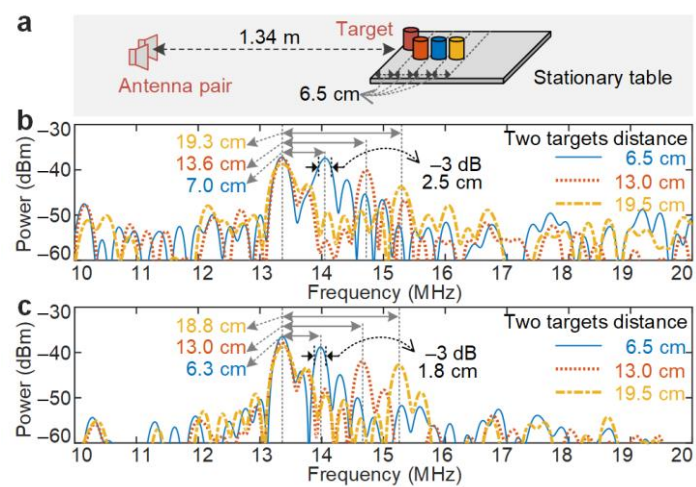

Fig. 7 Stationary target ranging results. **a** Schematic for two stationary targets ranging. Ranging results when the baud rate of the ASK-LFM signal is **b** 0.5 Gbaud and **c** 2 Gbaud.

Figure 7a shows the schematic of ranging for two stationary targets. In this study, the turntable no longer rotates, and two cylinders are used as the targets. The range between cylinder 1 and the antenna pair is 1.34 m, while that between cylinder 2 and the antenna pair is set to 1.405 m, 1.470 m, and 1.535m, respectively. Figure 7b and c show the ranging results, i.e., the electrical spectra after de-chirping, using the ASK-LFM signal with baud rates of 0.5 Gbaud and 2 Gbaud, respectively. From the frequency interval of the two strongest frequency components, the range between the two targets can be obtained, which is shown in Fig. 7b and c, with a maximum deviation from the actual value of only 0.7 cm. In addition, the 3-dB bandwidth of the peak after de-chirping is about 0.25 MHz (2.5 cm), and a good range resolution can be observed.

## 3.5 Spectrum sensing.

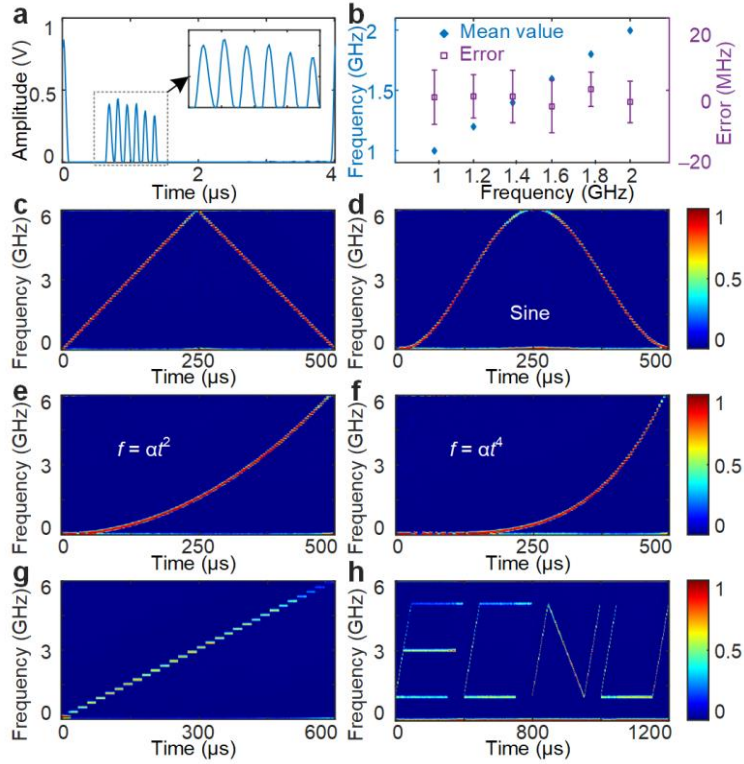

Fig. 8 Frequency and time-frequency measurement results. Frequency measurement results of six tones, **a** waveforms of the signal from PD1, **b** measurement error of 100 measurements. Measured time-frequency diagrams of different SUTs, **c** triangular-chirp LFM signal, **d** signal with "Sine" time-frequency diagram, **e** quadratic NLFM signal, **f** quaternary NLFM signal, **g** step-frequency signal, **h** signal with "ECNU" time-frequency diagram.

The spectrum sensing function is conducted using the same ASK-LFM signal. The LFM sideband after CS-TSSB modulation is the key to spectrum sensing, which has a bandwidth of 6 GHz and a sweep period of 4 μs. Because the BFS in the experiment is 10.8 GHz and the LFM sideband is 4.8 to 10.8 GHz away from the optical carrier, the spectrum sensing range is from 0 to 6 GHz. Figure 8a and b display the frequency measurement results for six tones from 1 to 2 GHz, with a frequency step of 0.2 GHz. Figure 8a exhibits the waveforms of the electrical signal obtained from PD1, which is captured by the OSC at a sampling rate of 100 MSa/s. Six distinct pulses corresponding to the six tones can be generated. The six frequencies are measured 100 times and the averaged values of the measured frequency are 0.9981, 1.1984, 1.3983, 1.5956, 1.8003, and 1.9968 GHz, respectively, which are shown in Fig. 8b. The standard deviations of the 100 measurements, represented by purple error bars in Fig. 8b, are 7.5, 6.0, 7.3, 7.3, 4.8, and 5.8 MHz. It is indicated that the frequency measurement error does not exceed ±10 MHz.

Figure 8c–h display the measured two-dimensional time-frequency diagrams of different SUTs via analog short-time Fourier transform using the proposed system. The diagrams in Fig. 8c–f clearly identify the triangular-chirp LFM signal, signal with a "Sine" time-frequency diagram, quadratic non-linear frequency-modulated (NLFM)

signal, and quaternary NLFM signal, with a signal duration of 500 μs. Figure 8g presents the measurement result of a step-frequency signal from 0.1 to 5.9 GHz with a frequency step of 200 MHz and a signal period of 600 μs. Additionally, Fig. 8h presents the measurement result of a signal with an "ECNU" time-frequency diagram, with a signal period of 1200 μs. "ECNU" is the abbreviation of East China Normal University. As can be seen, the time-frequency analysis for all these signals is presented with remarkable clarity.

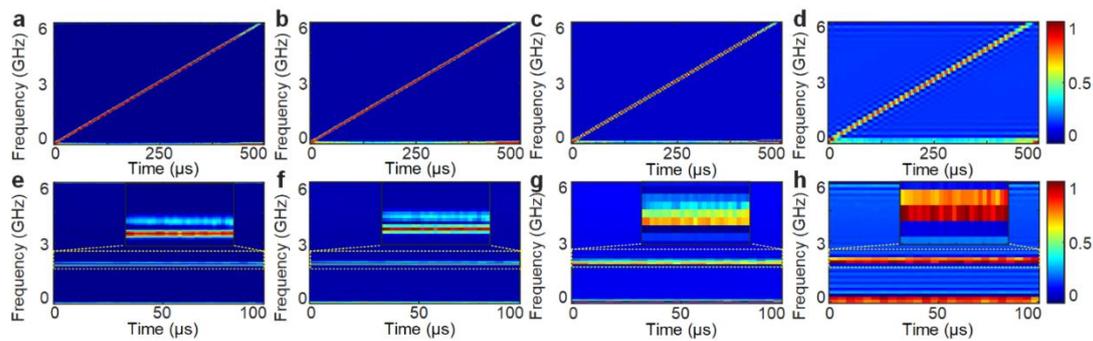

Fig. 9 Time-frequency diagrams at different sampling rates. Measured time-frequency diagrams of the LFM signal when the sampling rate is **a** 100, **b** 50, **c** 20, and **d** 10 MSa/s. Measured time-frequency diagrams of the two-tone signal with frequencies of 1.92 and 2 GHz processed at a sampling rate of **e** 100, **f** 50, **g** 20, and **h** 10 MSa/s.

In the previous experiment for spectrum sensing, the low-speed pulses from PD1 are sampled at a sampling rate of 100 MSa/s. To further investigate the spectrum sensing function at different sampling rates, more measurements are carried out. Figure 9a–d display the measured time-frequency diagrams of an LFM signal with a signal period of 500 μs at a sampling rate of 100, 50, 20, and 10 MSa/s. The results clearly indicate that the time-frequency diagrams exhibit a satisfactory frequency resolution when the processing sampling rate is 100 and 50 MSa/s. When the sampling rate is decreased to 20 MSa/s, there is a corresponding decrease in frequency resolution. When the processing sampling rate is further decreased to 10 MSa/s, the time-frequency diagram of an LFM signal exhibits a distinct step-frequency signal characteristic, which is caused by the reduced frequency resolution introduced by the low sampling rate.

To more clearly observe the influence of the sampling rate on the frequency resolution, an analysis is conducted on a two-tone signal with frequencies of 1.92 and 2 GHz at various processing sampling rates. Figure 9e–h depict the time-frequency diagrams of the two-tone signal at sampling rates of 100, 50, 20, and 10 MSa/s, respectively. As can be seen, when the sampling rate is 100 or 50 MHz, we can observe two separate lines in the time-frequency diagrams in Fig. 9e and f. However, as the sampling rate is further reduced to 20 MSa/s and 10 MSa/s, the two pulses for the two tones cannot be distinguished. Thus, in the time-frequency diagrams in Fig. 9g and h, the two lines touch each other, which means the frequency resolution of the spectrum sensing is reduced.

## 4. Discussion

As shown in Fig. 2, the frequency of the ASK-LFM signal can be adjusted by changing the center frequency of the IF ASK signal or the LFM signal applied to the DP-MZM. In the above experiment, the frequency of the ASK-LFM signal is mainly limited by the working frequency range of the antennas (8 to 18 GHz) and the EA (5.85 to 14.5 GHz) used in the RF link. Thus, in the experiment, the LFM signal is designed to have a frequency-sweep bandwidth from 4.8 to 10.8 GHz and the IF ASK signal is set with a center frequency at 3 GHz. Consequently, the ASK-LFM signal has a center frequency of 10.8 GHz in Fig. 3d and the subsequent experimental results.

Because the frequency measurement range of the spectrum sensing function is mainly determined by the frequency-sweep range of the LFM optical sideband and the position of the SBS gain spectrum, the frequency change of the LFM signal will affect the measurement range of the spectrum sensing function. Therefore, if only the center frequency of the electrical ASK-LFM signal needs to be adjusted, it is recommended to adjust the center frequency of the ASK-LFM signal. In the experiment, only a binary ASK signal is employed, which is mainly limited by the uneven frequency response of the mixer used in self-mixing and the low signal level at the receiving antenna. If mixers with a flat response or envelop detector with enough bandwidth are used in addition to greater receiving power, the system is also suitable for multi-level ASK signals.

In addition, the bandwidth of the ASK-LFM signal is jointly determined by the bandwidth of the IF ASK signal and the LFM signal. For different applications, the bandwidth of the ASK signal or LFM signal can be adjusted adaptively to achieve data communication at different data rates or target sensing with different precision. When the system bandwidth is limited, the implementation of the two functions needs to be compromised. When the bandwidth of the IF ASK signal or LFM signal is reduced to 0, the ASK-LFM signal is degraded to the LFM signal or ASK signal, respectively.

In the experiment, the optical carrier from the LD is directly used as the pump wave. Based on the above LFM signal settings, the measurement range of the spectrum sensing function is from 0 to 6 GHz in the experiment. The spectrum sensing bandwidth is limited by the bandwidth of the LFM signal, which can be easily changed by tuning the LFM signal bandwidth. The measurement range of the spectrum sensing function is determined by the relative position of the LFM optical sideband and the pump wavelength. To change the measurement range of the spectrum sensing function, two methods can be used: (1) Tuning the frequency-sweep range of the LFM signal; (2) Adjusting the position of the SBS gain spectrum, i.e., the pump wavelength. In order not to affect the generation of ASK-LFM signals, a more flexible way is the second one.

To adjust the pump wavelength, another DP-MZM (DP-MZM2) as a frequency shift module is added to the pump link immediately after the LD, as shown in Fig. 10a. Figure 10b shows the schematic diagrams of the signals at locations A–D in the system diagram. An RF signal from the microwave signal generator (MSG) is applied to the added DP-MZM, where CS-SSB modulation is implemented to frequency shift the optical carrier from the LD, as shown in Fig. 10b B. When the frequency of the RF signal from the MSG is set to $f_x$, the measurement range of the spectrum sensing function is changed from 0 to 6 GHz to from $f_x$ to $f_x$ + 6 GHz if the setting of the LFM signal is not changed. Figure 10b C-i and C-ii show two cases with a measurement range from 0 to $f_B$ when the RF frequency is 0 and a measurement range from $f_x$ to $f_B$ + $f_x$ when the RF frequency is $f_x$, respectively. It is worth noting that although the reference signal is used to determine the relationship between time domain pulses and the frequency information more conveniently in the experiment, the reference can be avoided by synchronizing the LFM signal transmission with the OSC acquisition by using a trigger signal.

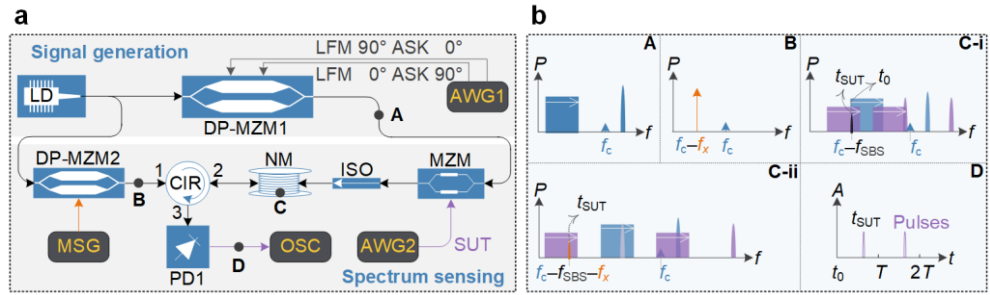

Fig. 10 Schematic of the proposed system. **a** System structure diagram that has an additional optical frequency shift module, i.e., DPMZM2, for tuning the measurement range of the spectrum sensing function. **b** Spectra and waveforms at different locations (A–D) in **a**.

## 5. Conclusion

In summary, we have introduced a photonics-enabled approach that seamlessly integrates radar ranging/imaging, wireless communications, and spectrum sensing within a unified framework. The key of the proposed approach is to generate an optical signal that can be employed to perform microwave ranging/imaging, and spectrum sensing in addition to wireless communication, which is done here using photonics to generate an ASK communication sideband and a linearly frequency-modulated (LFM) sideband, in which the LFM sideband is used for spectrum sensing, and the combination of the LFM and ASK sidebands is used for radar ranging/imaging and wireless communication functions. The proposed approach is evaluated experimentally. A joint microwave ranging/imaging, wireless communications, and spectrum sensing system supporting radar ranging with a measurement error within ±4 cm, two-dimensional imaging with a resolution of 25×24.7 mm, wireless communications with a data rate of 2 Gbaud, and spectrum sensing with a frequency measurement error within ±10 MHz in a 6-GHz bandwidth, is demonstrated. This research holds great significance for the 6G technology, enabling precise perception of the surrounding physical and electromagnetic environments while maintaining high-speed data communication.

## Acknowledgements
This work was supported by the National Natural Science Foundation of China under Grant 62371191 and Grant 61971193, the Natural Sciences and Engineering Research Council of Canada (NSERC), and the Science and Technology Commission of Shanghai Municipality under Grant 22DZ2229004.